\documentclass[letterpaper, 10 pt, conference]{ieeeconf}  

\IEEEoverridecommandlockouts                              

\usepackage{amsmath} 
\usepackage{amssymb}  
\usepackage[english]{babel}
\usepackage{lipsum}
\usepackage{nicematrix}
\usepackage{graphicx}
\usepackage{amsmath}
\usepackage{textcomp}
\usepackage{amssymb}
\usepackage{bm}
\usepackage{mathtools}

\usepackage{comment}
\usepackage{textcomp}
\usepackage{microtype}
\usepackage{float}
\usepackage{booktabs,makecell,tabularx}

\newcolumntype{C}[1]{>{\centering\arraybackslash}p{#1}}
\newcolumntype{L}{>{\raggedright\arraybackslash}X}

\usepackage{siunitx}
\usepackage{booktabs}
\usepackage{algorithm}
\usepackage{algpseudocode}
\usepackage{etoolbox}

\makeatletter
\newcommand*{\algrule}[1][\algorithmicindent]{%
  \makebox[#1][l]{%
    \hspace*{.2em}
    \vrule height .75\baselineskip depth .25\baselineskip
  }
}

\newcount\ALG@printindent@tempcnta
\def\ALG@printindent{%
    \ifnum \theALG@nested>0
    \ifx\ALG@text\ALG@x@notext
    \else
    \unskip
    \ALG@printindent@tempcnta=1
    \loop
    \algrule[\csname ALG@ind@\the\ALG@printindent@tempcnta\endcsname]%
    \advance \ALG@printindent@tempcnta 1
    \ifnum \ALG@printindent@tempcnta<\numexpr\theALG@nested+1\relax
    \repeat
    \fi
    \fi
}
\patchcmd{\ALG@doentity}{\noindent\hskip\ALG@tlm}{\ALG@printindent}{}{\errmessage{failed to patch}}
\patchcmd{\ALG@doentity}{\item[]\nointerlineskip}{}{}{} 
\makeatother

\newrobustcmd{\B}{\bfseries}
\title{\LARGE \bf
Voltage Restoration in MVDC Shipboard Microgrids with Economic Nonlinear Model Predictive Control
}

\author{Saskia Putri$^{1}$, Ali Hosseinipour$^{2}$,  Xiaoyu Ge$^{3}$, Faegheh Moazeni$^{4}$, Javad Khazaei$^{5,\ast}$
\\
\textcolor{blue}{This work has been submitted to the IEEE for possible publication. }\\ \textcolor{blue}{Copyright may be transferred without notice, after which this version may no longer be accessible.}
\thanks{This work was in part under support from the Department of Defense, Office of Naval Research award number N00014-23-1-2602. The authors are with the Rossin college of engineering and applied science at Lehigh University, Bethlehem, PA 18015, USA. (Emails: 
        {\tt\small sap322@lehigh.edu, alh621@lehigh.edu, xig620@lehigh.edu, moazeni@lehigh.edu,  khazaei@lehigh.edu)}}%
\thanks{$\ast$Corresponding author: Javad Khazaei. }
}

\begin{document}

\pdfminorversion=5 
\pdfcompresslevel=9
\pdfobjcompresslevel=4
\maketitle
\thispagestyle{empty}
\pagestyle{empty}

\begin{abstract}
Future Naval Microgrids (MGs) will include hybrid energy storage systems (ESS), including battery and supercapacitors to respond to emerging constant power loads (CPLs) and fluctuating pulsed power loads (PPLs). Voltage regulation of naval microgrids and power sharing among these resources become critical for success of a mission. This paper presents a novel control strategy using nonlinear model predictive controller embedded with a complex droop control architecture for voltage restoration and power sharing in medium voltage DC (MVDC) Naval MGs. The complex droop control ensures allocating supercapacitors (SCs) for high-frequency loads (i.e., PPLs), while battery energy storage system (BESS) and auxiliary generators share the steady-state load (i.e., CPL).  Compared to state-of-the-art control of the naval ship MGs that relies on linear models, the proposed method incorporates the nonlinear behavior of the MGs in the closed-loop control framework via nonlinear model predictive control (NMPC). A reduced order representation of the MVDC dynamic is employed as the prediction model, augmented with a multi-objective, constraints-based, optimal control formulation. The results demonstrate the effectiveness of the proposed control framework for voltage restoration and power sharing of resources in naval MGs.  
\end{abstract}
\section{INTRODUCTION}
With the rapid advancements in the technology and operation of navy ships, there is an increasing demand for the development of reliable and adaptable naval microgrid architectures \cite{cupelli2018voltage}. Traditional AC systems, characterized by their rigid architectures and the need for multiple power transformers that put a burden on the cargo weight, may not be adequate to accommodate future shipboard power systems \cite{faddel2019coordination}. Therefore, future ship electrification has moved towards medium voltage DC (MVDC) power systems \cite{chen2019novel,ieee2010ieee}. It offers a reliable, low-noise, and highly efficient power system equipped with environmental protection \cite{ieee2010ieee}. However, within MVDC power systems, formidable challenges arise, particularly in addressing fast, intermittent, and high-frequency power demands, such as those from pulsed power loads (PPL) during railgun operations, laser usage, and high-power radar application \cite{young2023model}. Furthermore, MVDC systems in naval vessels face issues pertaining to voltage variation and power quality degradation due to PPLs and continuous variation in propulsion speed \cite{khazaei2021optimal}. Consequently, an optimal control approach that is capable of effectively managing these multi-variate and complex systems is essential. 
\par Many studies have been conducted to optimize MVDC ship microgrids. In our preliminary results in \cite{khazaei2021optimal}, 
 we proposed a novel power flow optimization for MVDC shipboard power systems with hybrid energy storage systems. Optimal power flow was guaranteed using virtual resistive and capacitive controllers to manage power fluctuations and minimize operational costs subject to various system constraints. Moreover, energy management of MVDC power system was studied in \cite{nguyen2021energy} by optimizing load operability and ramp-rate characteristics of the energy storage systems (ESSs). The authors utilized receding horizon optimization to reduce the computational burden when using mixed-integer linear programming (MILP). However, these studies are performed in an open-loop configuration. A closed-loop framework that continuously allows correction of the control input based on the system's output is vital when optimizing the MVDC navy MGs. In \cite{bosich2014voltage}, the authors implemented a voltage control to accommodate CPLs in a ship under three control strategies: state feedback, active damping, and linearization via state feedback. The study, however, did not accommodate PPLs that pertain to being the most challenging load to supply due to their fast-changing demands within seconds. In addition, the control approach did not explicitly include all constraints related to generation units and voltage regulation. Given the existing challenges, model predictive control (MPC) has emerged as a powerful tool capable of handling constraints and providing a preview of the system output to yield an optimal control approach tailored to MVDC power systems characteristics \cite{bordons_model_2020}.  

MPC utilizes dynamical models of the system as the prediction model to optimize future control actions while accommodating system disturbances \cite{maciejowski2002predictive}. It solves online multiple open-loop control problems over a receding time horizon, subject to constraints. Various studies have been conducted on the energy management system of MVDC shipboard power systems using MPC. In \cite{zohrabi2017reconfiguration}, MPC is utilized to optimize power distribution to loads based on load priorities and operational constraints. However, the study did not consider voltage variation that may be caused due to the abrupt changes in the load and may continuously degrade the power system performance when left uncontrolled. In another study, the energy management system of navy MGs integrated with ESSs during high-ramp conditions was proposed \cite{van2017predictive}. Nonetheless, the study utilized a linear prediction model within the optimal control problem formulation of MPC. Although the trajectory of the prediction model showed similar behavior, a distinct level of the variables was observed. This may be due to the linear prediction model, which may suffer from a lack of representation of an actual system that exhibits nonlinear behavior. In addition, none of the existing MPC approaches considered power sharing between multiple energy resources, including batteries, supercapacitors, and generators within a ship. There is currently a research gap for a closed-loop energy management system of a navy ship that not only can regulate the DC voltage on the ship but also accommodate multiple energy resources in the presence of PPLs and CPLs. To address the aforementioned knowledge gaps, this paper proposes a novel NPMC formulation with complex droop control for power sharing among multiple resources and voltage regulation of MVDC shipboard MGs integrated with hybrid energy storage systems. The main contributions of the paper are:
\begin{enumerate}
   \item Voltage control algorithm of MVDC navy shipboard MGs is formulated using nonlinear (NMPC). The control algorithm is capable of restoring the voltage at a steady level with reliable power management to meet the pulsed and constant power loads.
   \item A complex power sharing algorithm is embedded in the MPC design to share the high-frequency PPL demand via supercapacitors and the steady-state CPL demand via generators and batteries.
    \item A reduced order model of MVDC Naval ship MGs is integrated with virtual capacitive and resistive droop controllers for energy resources within MG to ensure voltage restoration and power balance. 
    \item Optimal control formulation that provides a trade-off between power-rated and economically-driven power sharing between the energy resources is provided.   
    \item The impact of varying duration and magnitude of pulsed power load on the voltage stabilization performance is investigated. The control performance is compared to droop-controlled power management at the component level to validate the robustness and reliability of the proposed controller. 
\end{enumerate}

The rest of the paper is structured as follows. The architecture of the proposed MVDC is explained in Section~\ref{sec:MVDC} while the optimal control problem formulation from the NMPC framework is described in Section~\ref{sec:NMPC}. Section~\ref{sec:results} illustrates the validation of the proposed control algorithm through a variety of case studies. The conclusions are drawn in Section~\ref{sec:conc}. 
     \begin{figure}[t]
      \centering
      \includegraphics[width = 0.9\columnwidth]{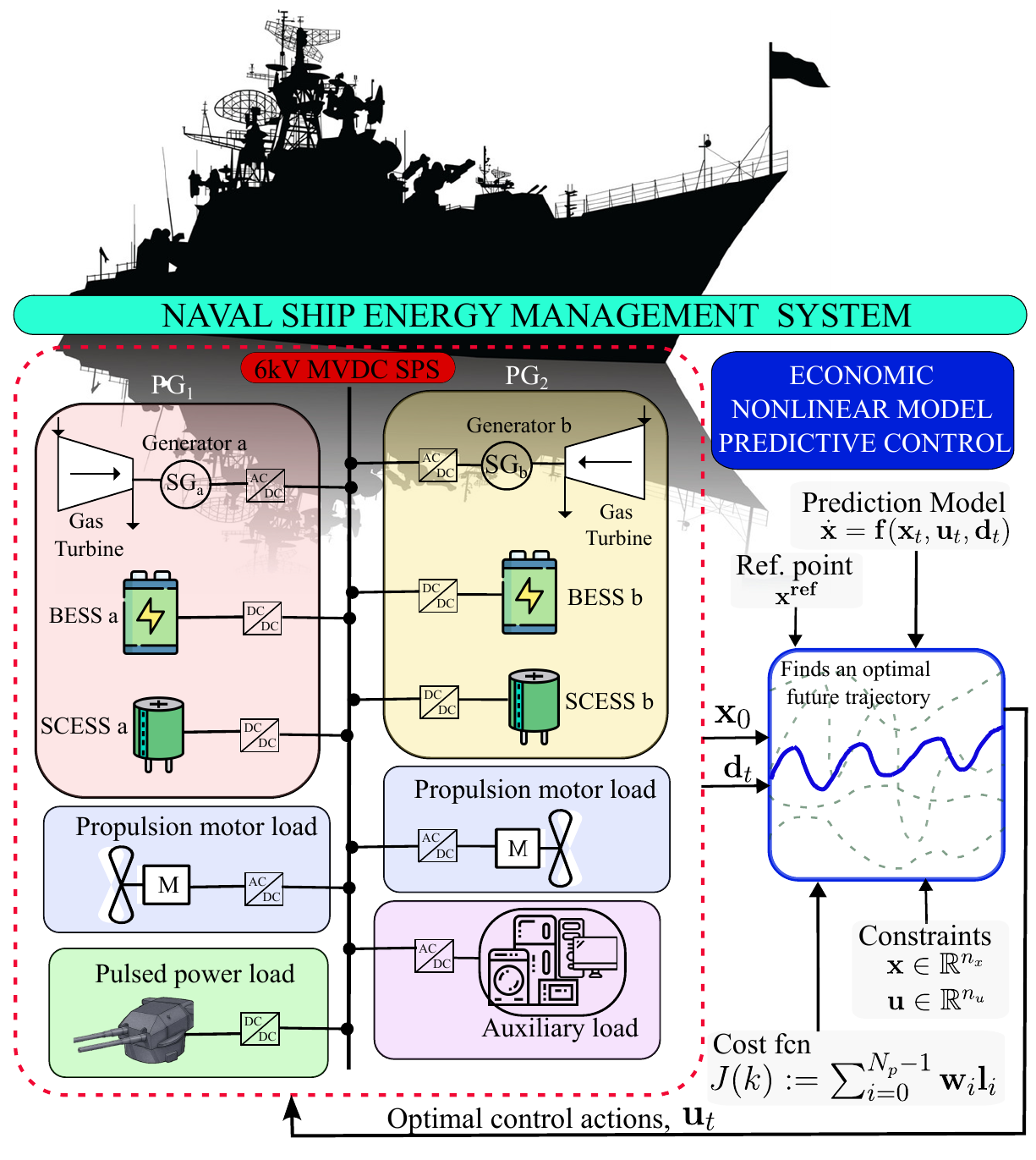}
      \caption{Proposed layout of the management of the MVDC shipboard power  system (SPS)}
       \vspace{-0.8cm}
      \label{mvdc}
   \end{figure}
\section{MVDC NAVY SHIPBOARD MICROGRIDS} \label{sec:MVDC}
\subsection{Proposed model description}
Fig.~\ref{mvdc} illustrates a generalized architecture of a 6kV shipboard power system. The generation unit in this system ensembles a pair of gas turbine-driven generators (SG), BESSs, and SCs. Considering the high energy density attributes of BESSs, they serve as backup sources for the SGs, ensuring a consistent power supply to meet the auxiliary and the propulsion motor loads (CPL). On the contrary, the SCs, equipped with a high power density feature, are intended to solely accommodate high-frequency power transients in the system stemming from the PPLs. Given the voltage variation that occurs in the system, the main control objective in this paper is to stabilize the voltage trajectory while ensuring load balance and power sharing between the distributed generation units (DGUs). In addition, a trade-off to economically dispatch the DGUs is offered through the addition of a cost-effective objective. As depicted in~Fig.~\ref{mvdc}, a summary of optimal control problem formulation is sent to the NMPC framework to find the optimal future trajectory to operate the proposed MVDC navy shipboard MGs effectively. 

\subsection{Reduced order model representation of the MVDC shipboard microgrids}
\begin{figure}[tb]
      \centering \vspace{0.05in}
      \includegraphics[width = 0.85\columnwidth]{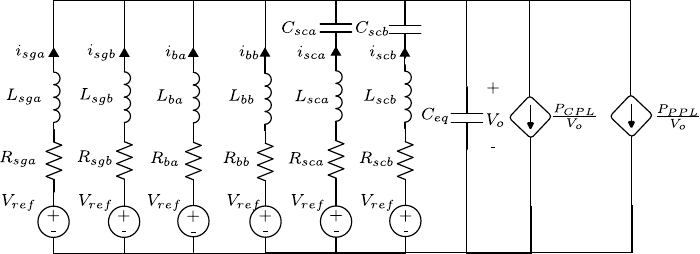}
      \caption{Equivalent circuit of the proposed MVDC shipboard microgrids}
      \vspace{-0.5cm}
      \label{circmvdc}
   \end{figure}

In this section, dynamic models of the proposed MVDC navy shipboard MGs are presented. Given the widespread practice of DC power systems with short distances, an equivalent circuit of the system can be derived that neglects cable longitudinal parameters \cite{sulligoi2014multiconverter}. Therefore, all capacitors and nonlinear loads can be designed in parallel, as illustrated in Fig.~\ref{circmvdc}. Extending from these assumptions, the dynamic models of the proposed system are expressed in~\eqref{dynmvdc}, developed from \cite{sulligoi2014multiconverter}.
\vspace{-0.1cm}
\begin{subequations} \label{dynmvdc}
\begin{align}
C_{\text{eq}} \dot{V_o} &= \frac{}{}\sum_{i \in \mathcal{N}_{SG}} (I_{\text{SG}i}) +\sum_{i \in \mathcal{N}_{B}} (I_{\text{B}i}) +\sum_{i \in \mathcal{N}_{SC}} (I_{\text{SC}i}) \nonumber \\
&- \frac{P_{\text{CPL}}}{V_o} -\frac{P_{\text{PPL}}}{V_o} \label{eqn:cap_voltage} \\
L_{i} \dot{I}_{i} &= V_{\text{ref}} - R_{i}I_{i} - V_o + \delta V_i \quad \forall i \in \{\mathcal{N}_{SG},\mathcal{N}_{B}\} \label{eqn:sg_current} \\
L_{i}\dot{I}_{i} &= V_{\text{ref}} - R_{i}I_{i} - V_{Ci} - V_o \quad \forall i \in \mathcal{N}_{SC} \label{eqn:sc_current} \\
C_{i}\dot{V}_{Ci} &= I_{i} \quad \forall i \in \mathcal{N}_{SC} \label{eqn:sc_voltage}
\end{align}
\end{subequations}
where $C_{eq}$ is the equivalent capacitor in $F$, $L_k$ is the inductance of the $k$-th generation unit in $H$, $V_o$ denotes the output voltage in $V$, $I_k$ as the current of the $k$-th generation unit in $A$, $R_{i}$, $C_{SCi}$ are the resistive droop gains for the conventional generators and BESSs, and capacitive droop gains for SCs, respectively. Details on the complex droop control design for the ship can be found in \cite{khazaei2021optimal,hosseinipour2023multifunctional}, and detailed parameters of the MGs are listed in Table~\ref{tab:mvdcparam}, adopted from \cite{hosseinipour2023multifunctional}. To ensure that the SCs are operated during transient conditions and deactivated during steady state, capacitive droop characteristics and voltages of the SCs are embedded in \eqref{eqn:sc_current}, with additional SCs' voltage dynamics in \eqref{eqn:sc_voltage} adopted from \cite{khazaei2021optimal,hosseinipour2023multifunctional}.
\subsection{State space model}\vspace{-0.2cm}
Accordingly, the state space representation of~\eqref{dynmvdc} can be summarized as follows:\vspace{-0.1cm}
\begin{equation}
    \dot{\bm{x}} = \mathbf{f}(\bm{x}(t),\bm{u}(t),\bm{d}(t)) \label{ssmvdc}\vspace{-0.2cm}
\end{equation}
where $\bm{x} \in \mathbb{R}^{n_x} = [V_o \ I_{SGa} \ I_{SGb} \ I_{Ba} \ I_{Bb} \ I_{SCa} \ I_{SCa}]^T$ is the state variable vector, $\bm{u}  \in \mathbb{R}^{n_u} = [\delta V_{SGa} \ \delta V_{SGb} \ \delta V_{Ba} \ \delta V_{Bb}]^T$ denotes the control input vector, and $\bm{d} \in \mathbb{R}^{n_d} = [P_{CPL} \ P_{PPL}]^T$ represents the disturbances of the system. In this paper, localized and centralized $\delta V$ will be used to offer different approaches to the power sharing. In the first case study, centralized $\delta V$ will be used with $u \in \mathbb{R}$, which guarantees power sharing based on the droop gains. In the second case study localized $\bm{\delta V}$ with $\bm{u} \in \mathbb{R}^{n_u}$ to ensure economic dispatch.

\begin{table}[tb!]
\centering
\caption{MVDC shipboard microgrids parameters} \vspace{-0.05in}
\label{tab:mvdcparam}
\resizebox{0.75\columnwidth}{!}{%
\begin{tabular}{@{}llll@{}}
\toprule
\textbf{Parameter} & \textbf{Qty} & \textbf{Parameter} & \textbf{Qty} \\ \midrule
$R_{SC_i}$ & 0.05 $\Omega$& $L_{SG_i}$ & 0.001 H\\
$R_{SG_a}$ & 0.05 $\Omega$& $L_{B_i}$ & 0.0008 H\\
$R_{SG_b}$ & 0.1 $\Omega$& $L_{SC_i}$ & 0.0004 H\\
$R_{B_a}$ & 0.225 $\Omega$& $C_{SC_a}$ & 5 F\\
$R_{B_b}$ & 0.45 $\Omega$& $C_{SC_b}$ & 10 F\\
$C_{eq}$ & 0.01 F&  &  \\
\midrule
\multicolumn{4}{c}{\textbf{NMPC simulation setup}} \\ \midrule
$T_{f}$ & 20 s & $V^{SP}$ & 6000 V \\
$t_s$ & 0.05 s & $\mathbf{Q}$ & 1 \\
$N_p$ & 10 & $\mathbf{R}$ & 0.001 \\ 
\midrule
\multicolumn{4}{c}{\textbf{CS-II: Cost-effective weight}} \\ \midrule
$\psi_{SG_i}$ & 0.002 & $\psi_{B_i}$ & 0.005 \\ 
$\psi_{SC_i}$ & 0.005 &\\
\bottomrule
\end{tabular}%
}
\vspace{-0.5cm}
\end{table}

\section{NONLINEAR MODEL PREDICTIVE CONTROL FRAMEWORK}\label{sec:NMPC}
MPC is comprised of four components: 1) a prediction model, 2) a set of constraints, 3) a cost function, and 4) an optimization algorithm \cite{bordons_model_2020}. In this work, the prediction model utilizes the discretized state-space model in \eqref{ssmvdc} over the prediction horizon ($N_p$), expressed as follows:
\begin{align}
    \bm{x}(k+j+1) &= \mathbf{f}(\bm{x}(k+j),\bm{u}(k+j),\bm{d}(k+j)) \nonumber \\ & \quad \quad \forall j \in \{0,\dots,N_p-1\} \label{nmpc_pred}
\end{align}
where $k = k_0 \to T -1$ represents the duration of the simulation time from the initial ($k_0$) to the final simulation time (T), $j$ is the sampling time-step over $N_p$. Note that in this paper, the prediction model fully utilizes the nonlinearity of the MGs in~\eqref{dynmvdc} to allow the controller to predict the system's future behavior accurately.

Furthermore, a set of constraints is explicitly added to account for the operable range of the bus voltage, currents of the energy sources, and the auxiliary voltage restoration signals assigned as the control inputs. Assuming full observability ($\mathbf{y}=\mathbf{x}$), the objective function is derived based on optimal control formulation over a finite horizon integrated with the cost-effective objective of the MVDC shipboard MGs. The optimization algorithm, incorporating the above-mentioned three components, is utilized to yield a sequence of optimal control actions over the prediction horizon, summarized as follows:  
\setlength{\abovedisplayskip}{2pt}
\setlength{\belowdisplayskip}{2pt}
\begin{subequations} \label{eq:ocp}
\begin{align}
    \min_{\mathbf{U}_k,\mathbf{X}_k} & J(k):=\sum_{j=0}^{N_p-1}\|\bm{x}(k+j)-\bm{x}^{ref}(k+j)\|^2_{\mathbf{Q}_j} \nonumber \\ &\quad +\|\bm{\Delta u}(k+j)\|^2_{\mathbf{R}_j} + \|\mathbf{x}(k+j)\|^2_{\mathbf{\Psi}_j} \label{c-objfun}\\
    \text{s.t.} \quad & \bm{x}(k+j+1) = \mathbf{f}(\bm{x}(k+j), \bm{u}(k+j),\bm{d}(k+j)) \nonumber \\
    & \quad \quad \quad \quad \quad  j = 0, 1, \dots, N_p-1  \label{c-dynamic}\\ 
    & \bm{u}(k+j) \in \mathcal{U}, \quad j = 0, 1, \dots, N_p-1 \label{c-ubound}\\ 
    & \bm{x}(k+j) \in \mathcal{X}, \quad j = 1, 2, \dots, N_p-1 \label{c-statebound}\\
    & \bm{\tilde{x}}_0 = \bm{x}_{plant}(t_0) \label{c-plant}
\end{align}
\end{subequations}
where $N_p \in \mathbb{N}^+$ is the prediction horizon, $k:= kt_s \in \mathbb{R}^+$ is the current time step at sampling time $t_s$, $\mathbf{Q} \succ 0$, $\mathbf{\Psi}\succ 0$, and $\mathbf{R}\succ 0$ are the penalty weights for the states and the rate control inputs, respectively. Also, $\bm{\Delta u}(k) \in \mathbb{R}^{n_u}$ denotes the input's rate of change vector at time step $t$, $\bm{x}^{ref} \in \mathbb{R}^{n_x}$ is the vector of desired set-points, and \eqref{c-dynamic} is the discretized prediction model. The first term of the objective function in~\eqref{c-objfun} ensures the setpoint tracking such that the bus voltage remains at a constant level, while the second term limits the step changes of the restoration signals. Given that the voltage variation is restricted to within $\pm 4\%$ with $I_i \approx P_i$, the cost associated with using the DGUs ($\bm{\psi} \in \mathbb{R}^{n}$) is added as the third term in \eqref{c-objfun}, thereby, achieving economic dispatch. Note that the third term is only applied to the second case study. Table~\ref{tab:mvdcparam} detailed the control setup for utilizing NMPC. An overview of NMPC implementation in this work can be found in \textbf{Algorithm~\ref{alg:nmpc}}.
\begin{algorithm}[b]
\caption{Economic Nonlinear MPC(NMPC)}
\label{alg:nmpc}
\begin{algorithmic}[1]
\State \textbf{Input Parameters:} $N_p,\mathbf{Q},\mathbf{R},\mathbf{\Psi},T_f,n_x,n_u$
\State \textbf{Input Data:} $\mathbf{x}^{ref}, \mathbf{f}(\bm{x}(t),\bm{u}(t),\bm{d}(t))$
\State \textbf{Initialize} $\bm{x}_0 \in \mathbb{R}^{n_x},\bm{u}_0 \in \mathbb{R}^{n_u}, \bm{d}_0 \in \mathbb{R}^{n_d},k=0$
\For {$\mathrm{k=0 \to T_f-1}$} \Comment{Simulation time}
\For {$j=0$ to $N_p-1$}
\State \textbf{Define \& Discretize} state-space model, \eqref{ssmvdc}
\State \textbf{Construct} prediction model, \eqref{nmpc_pred}
\State \textbf{Formulate} $J(\bm{x}(k+j),\bm{u}(k+j))$, \eqref{c-objfun}
\EndFor
\State \textbf{Solve} $J(k+j)$ s.t. \eqref{c-dynamic}-\eqref{c-statebound}
\State \textbf{Extract} $[{\bm{u}^{\ast}(0|k),\hdots,\bm{u}^{\ast}{(N_p-1|k)}}]$.
\State \textbf{Apply} only $\bm{u}^{\ast}(0|k)$ \Comment{RHC}
\State \textbf{Measure} $\bm{x}(k+1|k)$ from \eqref{dynmvdc}
\State \textbf{Update} for $k+1$, $\bm{x}_0 = \bm{x}(k+1|k)$ and $\bm{u}_0 = \bm{u}^{\ast}(0|k)$ 
\EndFor
\end{algorithmic}
\end{algorithm}
\begin{figure}[tb!]
      \centering
      \includegraphics[width =0.8 \columnwidth]{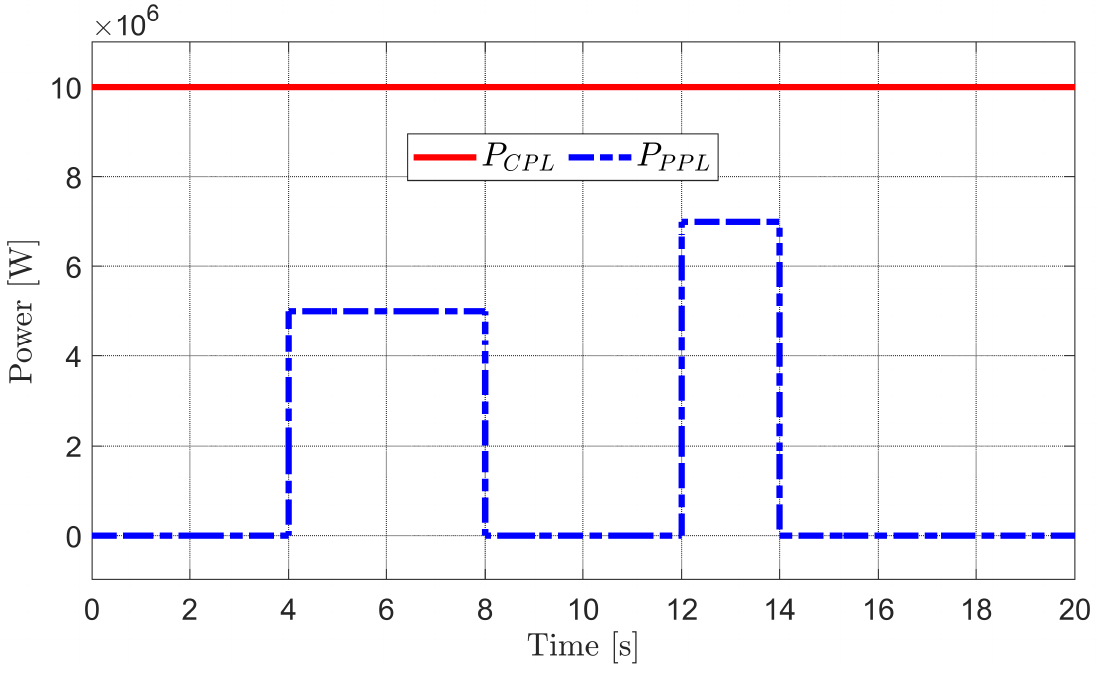}
      \caption{Load profile of the MVDC shipboard MGs with onboard pulsed power loads (PPL) and constant power loads (CPL).}
      \vspace{-0.5cm}
      \label{cs1:load}
   \end{figure}
 \begin{figure}[tb!]
      \centering
      \includegraphics[width = 0.8\columnwidth]{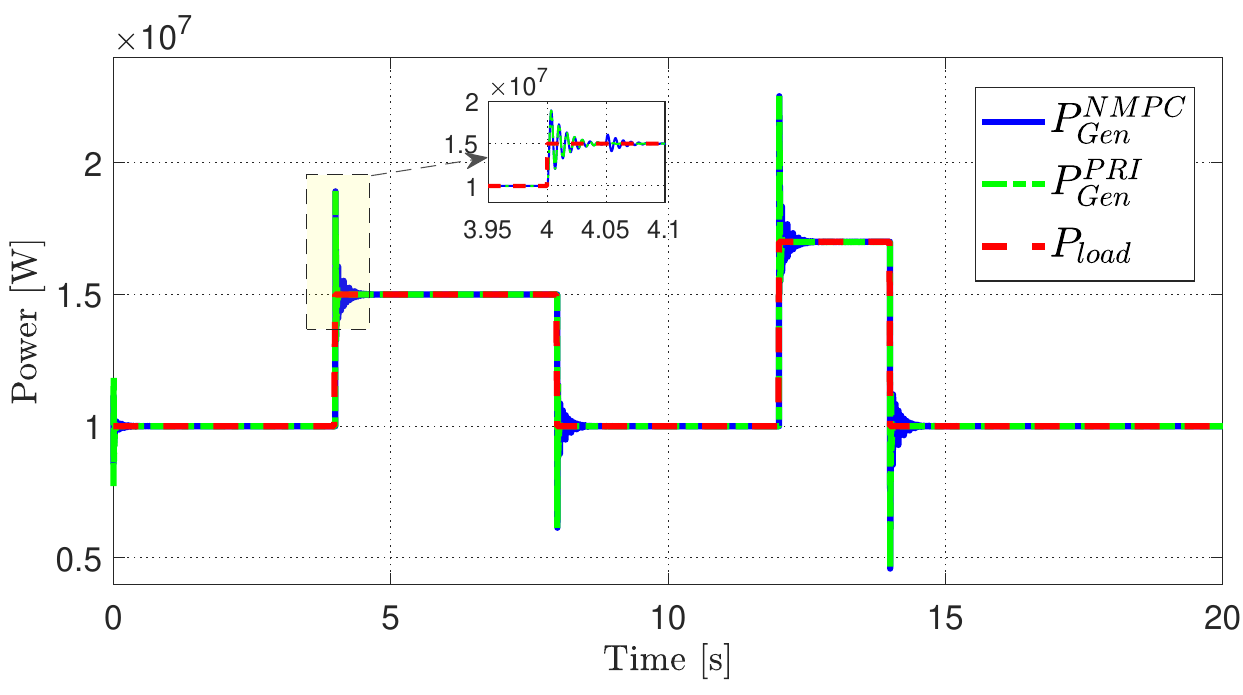}
      \caption{Power balance of the MVDC shipboard MGs under primary control and $\delta V^c$-NMPC integration.}
      \vspace{-0.5cm}
      \label{cs1:pwrgen}
\end{figure}
\section{CASE STUDIES} \label{sec:results}
The nonlinear optimization problem in~\eqref{eq:ocp} is solved in MATLAB using ``fmincon" nonlinear solver with a solution method specified to sequential quadratic programming (SQP). The models are carried out on an Intel Core CPU i7-6700 processor at 3.40 GHz and 32GB RAM. Two case studies are performed in this paper to verify the effectiveness of the proposed nonlinear MPC algorithm for the MVDC navy shipboard MGs. The main objective is to maintain the nominal voltage of the MVDC bus and confirm the load-power generation balance. A trade-off between power sharing at a component level and power allocation based on cost-effectiveness is provided in the first and second case studies, respectively. Mean absolute percentage error (MAPE) is used to quantify the performance, as follows: 
\begin{equation}
MAPE = \frac{1}{n} \sum_{i=1}^{n} \left| \frac{{x_i - \hat{x}_i}}{{x_i}} \right| \times 100\%  \label{mapeeq}
\end{equation}
where, $n$ represents the total number of samples, $x_i$ represents the desired value, and $\hat{x}_i$ represents the predicted value.

A consistent pattern of the load profile is utilized in all case studies, as illustrated in Fig.~\ref{cs1:load}. 

     \begin{figure}[tb!]
      \centering
      \includegraphics[width = 0.85\columnwidth]{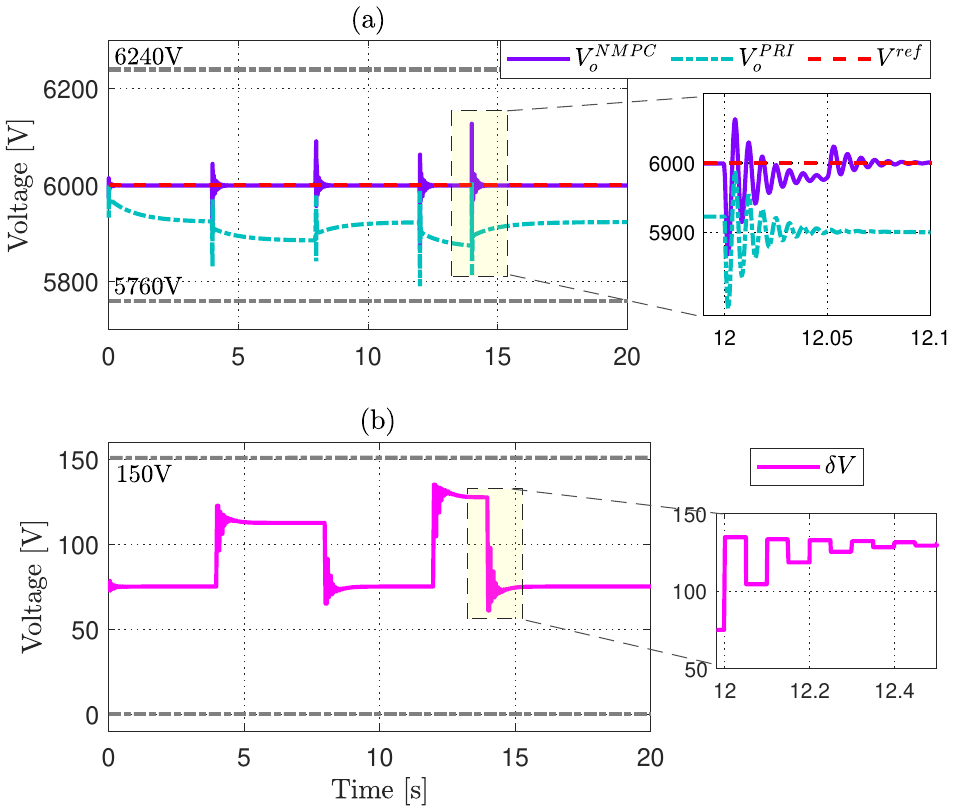}
      \caption{CS-I: Output voltage of the MVDC shipboard MG when using only primary control versus nonlinear MPC.}
      \vspace{-0.5cm}
      \label{cs1:volt}
   \end{figure}
        \begin{figure}[tb!]
      \centering
      \includegraphics[width = 0.85\columnwidth]{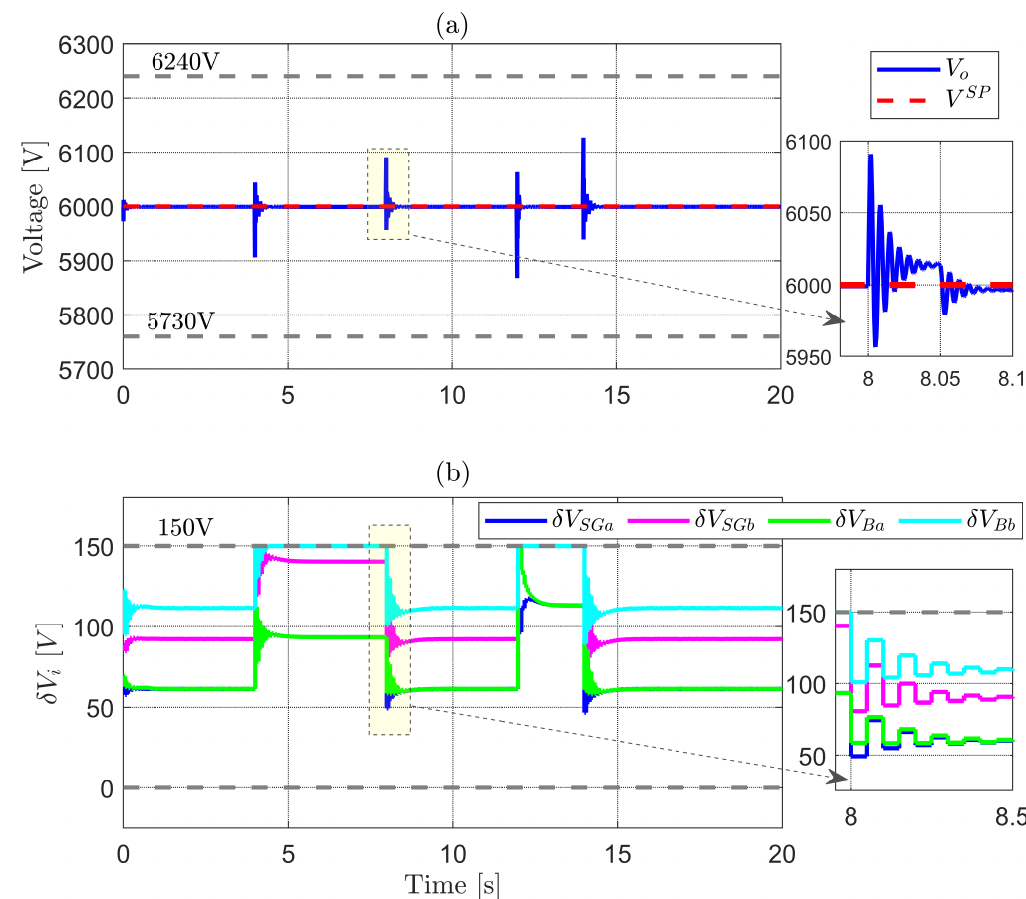}
      \caption{CS-II: Output voltage of the MVDC shipboard MG via economic NMPC.}
      \vspace{-0.5cm}
      \label{cs2:volt}
   \end{figure} 
      \begin{figure*}[t]
      \centering
      \includegraphics[width = 0.8\textwidth]{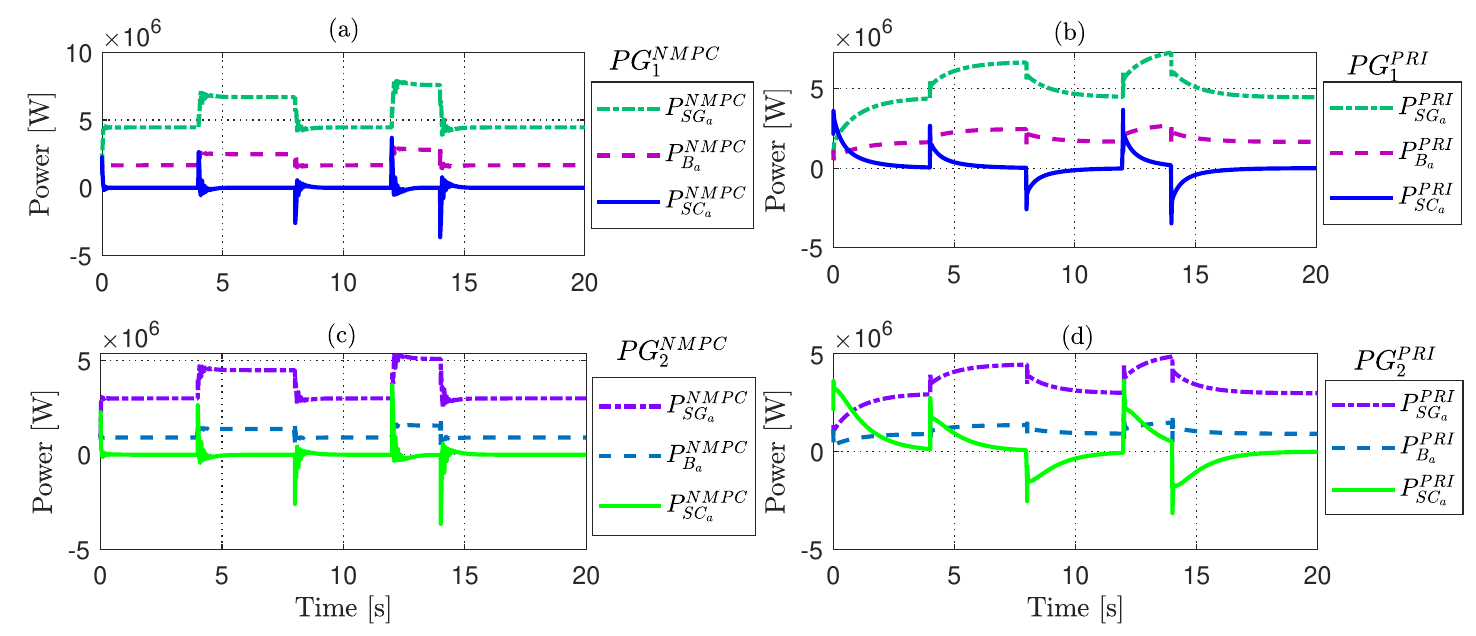} \vspace{-0.05in}
      \caption{Output power comparison of the distributed generation units in MVDC shipboard MGs.}
      \vspace{-0.5cm}
      \label{comparePRI}
   \end{figure*}
 \subsection{Case study 1 (CS-I): NMPC with centralized auxiliary voltage restoration signal ($\delta V^c$)}
 In this case study, a classical NMPC formulation is employed, prioritizing control objectives related to voltage stability and stable output power with proper load sharing.  To validate the efficacy of the integrated centralized voltage restoration signal using NMPC, herein referred to as $\delta V^c$-NMPC, its performance is collocated with a conventional droop-controlled power management system from the primary layer, herein referred to as the primary control. Fig.~\ref{cs1:pwrgen} shows the aggregate power generated from the DGUs against the total power demand for both control configurations ($P^{PRI}_{Gen}$ represents primary control via virtual droop control and $P^{NMPC}_{Gen}$ designates $\delta V$-NMPC). It is observed that both controllers consistently converge to the load trajectory, thereby ensuring adherence to the demand. 

Fig.~\ref{cs1:volt}(a) shows the voltage trajectory over the 20s simulation with both controllers. As expected, solely using virtual droop control is not sufficient to restore the voltage of the MVDC bus to the nominal value with MAPE at 1.67\%. This is due to the presence of voltage-sensitive loads, PPL, which required the system to withdraw the currents excessively. As a result continuous voltage drop is observed.  On the contrary, $\delta V^c$-NMPC outperforms the primary control with voltage consistently reaching the reference point at 6kV with MAPE at 0.02\%. A transient state of the voltage trajectory is also observed during the sudden changes in the loads. In the proposed method, the excessive withdrawal of the currents is managed by the bounded auxiliary voltage signal, exhibited in Fig.~\ref{cs1:volt}(b), ensuring voltage stabilization to the nominal value.  

In Fig.~\ref{comparePRI}, the power generation trajectories of the DGUs under both primary control and $\delta V^c$-NMPC integration are exhibited. In the presence of time-varying PPL, $\delta V^c$-NMPC managed to stabilize the power output of each DGU four times faster than the primary control, confirming the robustness of the proposed controller. Furthermore, the incorporation of capacitive droop gains into the dynamics of the SCs played a significant role in the power output, where they consistently adapt to transient load conditions. Under the $\delta V^c$-NMPC configuration, SCs are rapidly adjusted to 0MW during steady state conditions within 0.5s, whereas the primary control requires a 2s longer time frame to reach a similar condition. 

Fig.~\ref{cs1:pwrshare} shows the power sharing between the DGUs of the MVDC navy MG. During steady state conditions, the output power of $SG_a$ and $BESS_a$ approximately doubled that of $SG_b$ and $BESS_b$. This is because the droop gains (see Table~\ref{tab:mvdcparam}) are reversely proportional to the steady state power \cite{hosseinipour2023multifunctional}.

\subsection{Case study 2 (CS-II): Economic NMPC with localized auxiliary voltage restoration signals ($\delta V^l_i$)}
In this case study, the proposed economic NMPC (ENMPC) is utilized to also factor in the cost-effectiveness of the power allocation in the MVDC navy MG. The final optimization problem formulation is detailed in~\eqref{eq:ocp}. In addition, localized auxiliary voltage restoration signals ($\delta V_i$) for the $i$-th DG are employed to regulate the loss at the component level. Given the power balance constraint embedded in the system's dynamics, the total power generated by the MGs consistently complies with the total load (figure not shown for brevity) with a similar trajectory as shown in CS-I. During sudden load fluctuations, the system experiences noticeable voltage drops and surges as illustrated in Fig~\ref{cs2:volt}. However, the proposed controller maintained a voltage stability of less than 0.5s. 
The addition of $\delta V_i$ to the system's dynamics is used to compensate for the virtual impedance ($\delta V_i \approx R_iI_{i}$) such that constant MVDC bus voltage can be obtained. In both case studies, it is visualized that $\delta V$ fulfilled the objective. Furthermore, leveraging the MPC framework that can explicitly take into account constraints on the system's variables, here, $\delta V_i$ for each DGU is bounded to 150V to be consistent with the boundary in CS-I. As observed in Fig~\ref{cs2:volt} (b), optimal control actions are achieved that satisfy the constraints, secure power balance, and restore the bus voltage.  

\begin{figure}[tb!]
      \centering
      \includegraphics[width = 0.95\columnwidth]{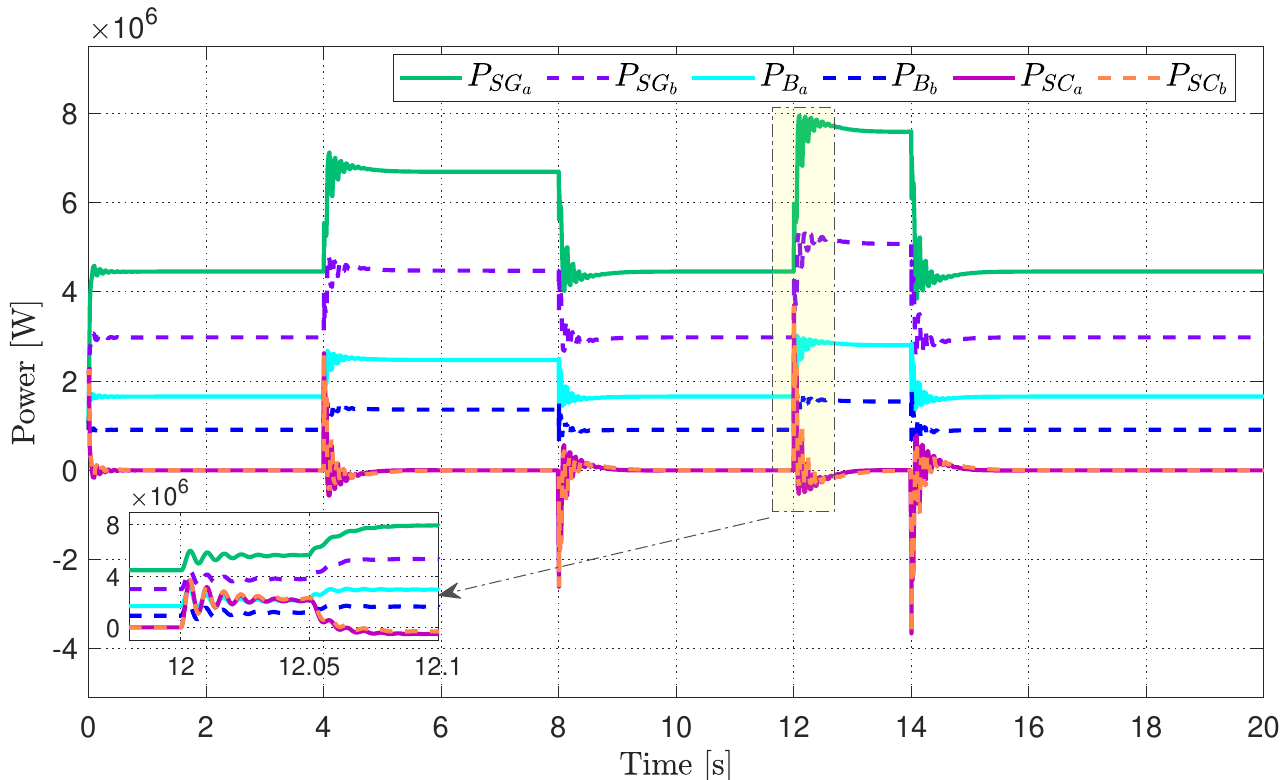}
      \caption{CS-I: Power sharing with $\delta^C$-NMPC.}
      \vspace{-0.2cm}
      \label{cs1:pwrshare}
   \end{figure}

\begin{figure}[t]
      \centering
      \includegraphics[width = 0.95\columnwidth]{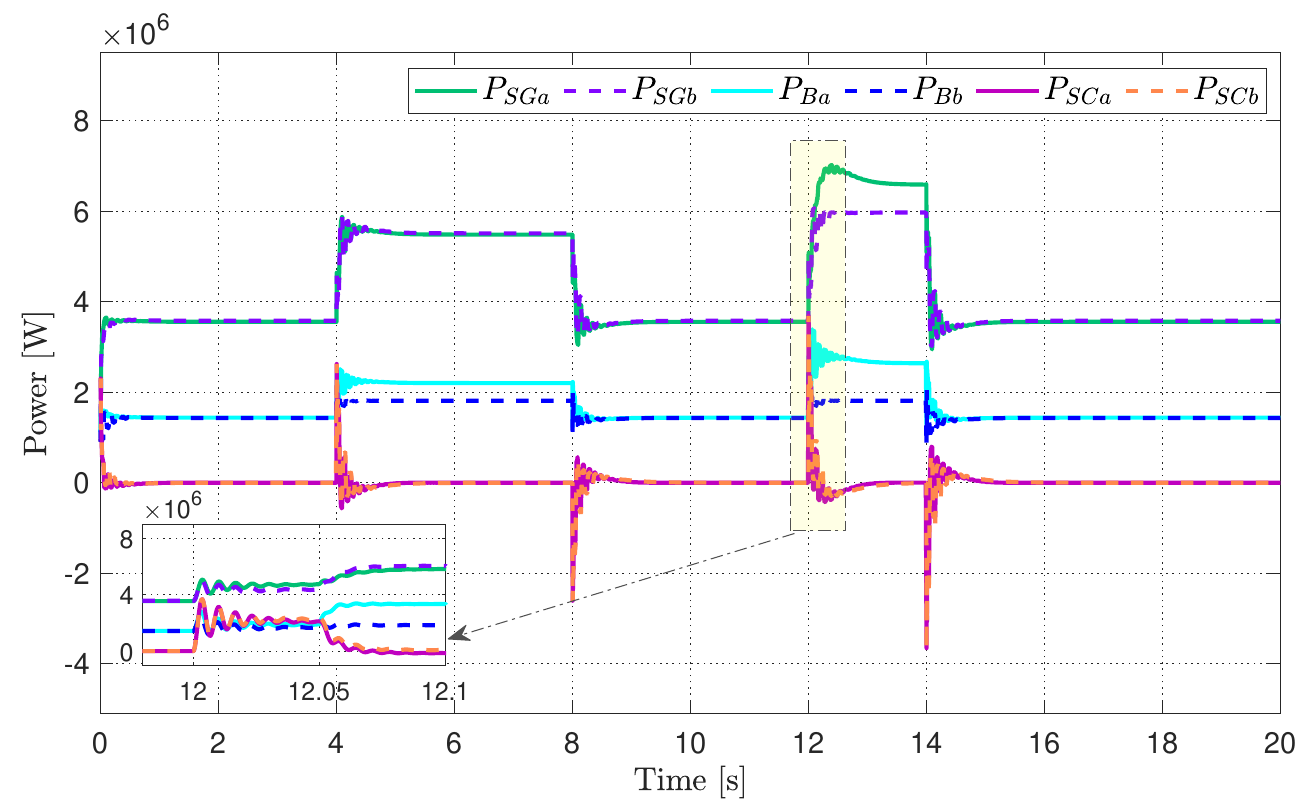} \vspace{-0.05in}
      \caption{CS-II: Power sharing with $\delta^l$-ENMPC.}
      \label{cs2:pshare}
      \vspace{-0.8cm}
 \end{figure}

Fig.~\ref{cs2:pshare} shows the load sharing among the DGUs. Generally, the order of the output power is consistent with CS-I during steady state conditions (see Fig.~\ref{cs1:pwrshare}), where $P_{SGa}$ contributes the highest power while $P_{Bb}$ is the lowest and $P_{SCi}$ only utilized during transient conditions. However, the ratio between the DGUs differs, closely reaching 1:1. Given the economic objectives in~\eqref{c-objfun}, this result is anticipated. When the objectives of the composite optimization conflict with one another, a compromised solution is generated in which no improvement can be made without deteriorating at least one other solution \cite{bordons_model_2020}. Furthermore, 15\% cost savings are obtained in this case study when compared to CS-I with cost coefficients adopted from \cite{roy2021economic, eia2013levelized}. 

\section{CONCLUSIONS} \label{sec:conc}
This paper proposes a novel approach for voltage control and power sharing among hybrid storage and conventional generation units in MVDC navy shipboard microgrids by leveraging a nonlinear model predictive control framework. 
The results demonstrate that both approaches constantly maintain the bus voltage at a steady level with a tracking error of 0.02\%, in spite of the varying, voltage-sensitive, and high load requirement. The controller adeptly stabilizes the generated power within 0.5s during abrupt changes in the load. These results confirm the robustness of the controller. Furthermore, demand compliance is effectively achieved across the DGUs with SGs and BESSs catering to the constant power loads ($P_{CPL})$, while SCs adeptly supply the demands from the pulsed power loads ($P_{PPL}$). 
Moreover, the integration of the MVDC navy shipboard MGs with NMPC as the controller allows a flexible tuning mechanism for power allocation. Droop-control-based power sharing is achieved when utilizing centralized-$\delta V$. In contrast, localized-$\delta V_i$ prioritizes economic power generation with 15\% cost savings. 
Future work will develop a fast and guaranteed stability trajectory of the MVDC MGs for naval vessels. The algorithm will leverage the Lyapunov stability theorem and input parameterization under different layers of hierarchical control in MGs. 

    \section*{ACKNOWLEDGMENT}
 The authors would like to acknowledge the Department of Defense, Office of Naval Research for supporting this research.  
\addtolength{\textheight}{-10cm}   





\vspace{-0.5cm}
\bibliographystyle{IEEEtran}
\bibliography{IEEEabrv,main}

\end{document}